\documentclass[preprint,showpacs,showkeys,preprintnumbers,aps]{revtex4}

\usepackage{graphicx}

\begin{document}

\preprint{MMM, Fecher, DS-04}

\title{Slater-Pauling Rule and Curie-Temperature
       of Co$_2$-based Heusler compounds.}

\author{Gerhard H. Fecher, Hem C. Kandpal, Sabine Wurmehl,
        and Claudia Felser}
\affiliation{Institut f\"ur anorganische und analytische Chemie,
             Johannes Gutenberg - Universit\"at,
             D-55099 Mainz, Germany}
\author{Gerd Sch\"onhense}
\affiliation{Institut f\"ur Physik,
             Johannes Gutenberg - Universit\"at,
             D-55099 Mainz, Germany}
\date{\today}

\begin{abstract}
A concept is presented serving to guide in the search for new
materials with high spin polarization. It is shown that the magnetic
moment of half-metallic ferromagnets can be calculated from the
generalized Slater-Pauling rule. Further, it was found empirically
that the Curie temperature of Co$_2$ based Heusler compounds can be
estimated from a seemingly linear dependence on the magnetic moment.
As a successful application of these simple rules, it was found that
Co$_2$FeSi is, actually, the half-metallic ferromagnet exhibiting
the highest magnetic moment and the highest Curie temperature
measured for a Heusler compound.
\end{abstract}


\maketitle

\section{Introduction}

There is a growing interest in materials with high spin
polarization. Half-metallic ferromagnets (HMF) seem to be the
materials of choice for applications, due to their exceptional
electronic structure. They are metals for one spin direction and
semiconductors for the other. This means that the electrons are
100\% spin polarized at the Fermi energy.

The present research concentrates on finding potential materials
with high spin polarization in the class of Heusler compounds. Their
half-metallicity with respect to the spin was predicted by de Groot
\cite{gme83} for half-Heusler and by Ishida \cite{imf98} for Heusler
compounds. The latter have the chemical formula $X_2YZ$ with $X$ and
$Y$ being transition metals and $Z$ being a main group element.
Heusler compounds crystallize in the $L2_1$ structure
($F\:m\overline{3}m$). In particular, the Co$_2$YZ compounds exhibit
the highest Curie temperature and the highest magnetic moments per
unit cell (see data in \cite{LB19C,LB32C}).

\section{Calculation of the electronic structure}

The electronic structure of most of the known ternary Heusler
compounds was calculated in order to find their magnetic moments and
magnetic type. The calculations were performed by means of the full
potential linear augmented plane wave (FLAPW) method as provided by
Wien2k \cite{bsm01}. The exchange correlation energy functional was
parameterized within the generalized gradient approximation
(GGA)\cite{pbe96}. The energy convergence criterion was set to
10$^{-5}$. For $k$-space integration, a $20\times20\times20$ mesh
was used resulting in 256 $k$ points of the irreducible part of the
Brillouin zone.

Overall, the calculations were performed for 59 Heusler compounds
based on $X_2$ and $Y$ being $3d$ metals, 17 with only $X_2$ and 28
with only $Y$ being a $3d$ metal, as well as some compounds
containing rare earth metals. In the first two groups, the heavy
$3d$ elements (Mn, Fe, Co, Ni, and Cu) were placed on X sites. In
the first group, the Y positions were occupied by Sc,...,Ni with the
restriction that the Y element was always lighter than the element
on X positions or equal. For the third group, the X position was
occupied by the $4d$ elements Ru, Rh, Pd, Ag or by the $5d$ elements
Ir, Pt, Au, and the Y position mostly by light $3d$ elements. It
turned out that nearly all (if not paramagnetic) Co based compounds
(Co$_2$YZ) should exhibit half-metallic ferromagnetism. The
calculated magnetic moments were used for an analysis by means of
the Slater-Pauling rule as described in the following section.

\subsection{Slater - Pauling rule for Heusler compounds}

Slater \cite{sla36} and Pauling \cite{pau38} reported first that the
magnetic moments ($m$) of $3d$ elements and their binary compounds
can be described by the mean number of valence electrons ($n_V$) per
atom. The rule distinguishes the dependence of $m(n_V)$ into two
regions. The first (closed packed structures: fcc, hcp) is the range
of itinerant magnetism ($n_V\geq8$) and the second (bcc) is the one
of localized moments ($n_V\leq8$), where Fe is a borderline case.
According to Hund's rule it is often favorable that the majority $d$
states are fully occupied ($n_{d\uparrow}=5$). Starting from $m = 2
n_{\uparrow} - n_V$, this leads to the definition of the magnetic
valence to be $n_M = 10 - n_V$ such that the magnetic moment per
atom is given by $m = n_M + 2 n_{sp\uparrow}$. Pauling gave a value
of $n_{sp\uparrow} \approx 0.3$ for the second region. A plot of $m$
versus magnetic valence ($m(n_M)$ is called the generalized
Slater-Pauling rule, as described by K\"ubler \cite{kue84}. In the
case of localized moments, the Fermi energy is pinned in a deep
valley of the minority electron density. This constrains
$n_{d\downarrow}$ to be approximately three with the result $m
\approx n_V - 6 - 2 n_{sp\downarrow}$, for Fe and its bcc-like
binary alloys (Fe-Cr, Fe-Mn, and partially Fe-Co). It was shown by
Malozemoff {\it et al}\cite{mwm84} using {\it band-gap theory} that
these arguments hold principally also if using more realistic band
structure models, even so they were initially derived from rigid
band models. In particular, they have shown that the rule is still
valid if metalloids are involved.

Half-metallic ferromagnets are supposed to exhibit a real gap in the
minority density of states where the Fermi energy is pinned. The gap
has the consequence that the number of occupied minority states has
to be an integer. Thus, the Slater-Pauling rule will be strictly
fulfilled with
\begin{equation}
       m_{HMF} = n_V - 6
\label{eq1}
\end{equation}
for the spin magnetic moment per atom.

For ordered compounds with different kind of atoms it may be more
convenient to use all atoms of the unit cell. In the case of 4 atoms
per unit cell, as in Heusler (H) compounds, one has to subtract 24
(6 times the number of atoms) from the accumulated number of valence
electrons $N_V$ ($s, d$ electrons for the transition metals and $s,
p$ electrons for the main group element) to find the spin magnetic
moment $M$ per unit cell:
\begin{equation}
       M_{H} = N_V - 24 .
\label{eq2}
\end{equation}
This {\it rule of thumb} is strictly fulfilled for HMF only as first
noted in \cite{kue84} for half-Heusler (hH) compounds ($M_{hH} = N_V
- 18)$. In both types of compounds (X$_2$YZ and XYZ) the spin
magnetic moment per unit cell becomes strictly integer for
half-metallic ferromagnets. An already very small deviation from an
integer value indicates that the HMF character is completely lost.
This situation changes for alloys with non-integer site occupancies
like the quaternaries X$_2$Y$_{1-x}$Y'$_x$Z. In such cases $M$ may
become non-integer depending on the composition, even for the HMF
state.

The Slater-Pauling rule relates the magnetic moment with the number
of valence electrons, but is not formulated to predict a
half-metallic ferromagnet. The gap in the minority states of Heusler
compounds or other HMF has to be explained by details of the
electronic structure (for examples see \cite{kws83,gdp02}).

\section{Results}
\begin{figure}
\centering
\includegraphics[width=2.8in]{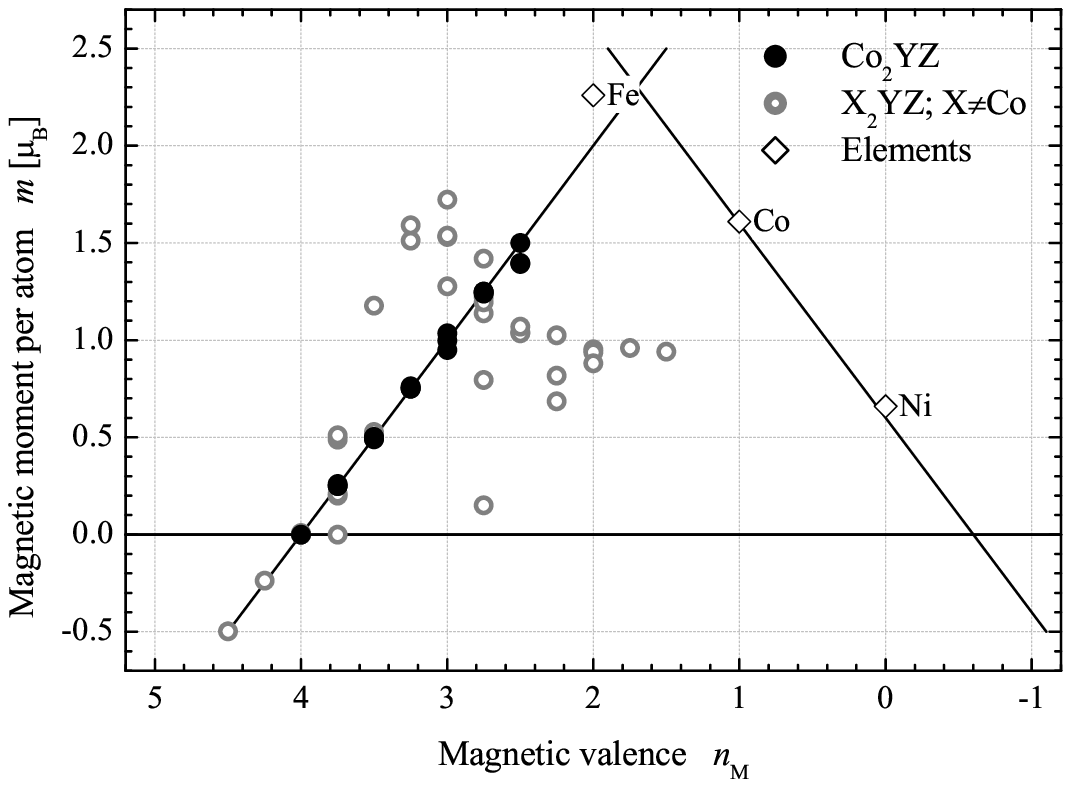}
\caption{Slater-Pauling graph for Heusler compounds. \\
         The Co$_2$ based Heusler compounds are marked by full dots.
         The elemental metals Fe, Co, and Ni are given for comparison.}
\label{fig1}
\end{figure}

Figure \ref{fig1} shows the generalized Slater-Pauling behavior of
selected Heusler compounds with $3d$ transition metals on the X and
Y sites in comparison to the magnetic elements Fe, Co, and Ni. The
magnetic moments were calculated as described above. It is seen that
the Co$_2$YZ compounds strictly fulfill the Slater-Pauling rule,
whereas other compounds exhibit pronounced deviations from the
Slater-Pauling like behavior. The latter, however, do not exhibit
half-metallic ferromagnetism.

Inspecting the other transition metal based compounds, one finds
that compounds with magnetic moments above the expected
Slater-Pauling value are X=Fe based. Those with lower values are
either X=Cu or X=Ni based, with the Ni based compounds exhibiting
higher moments compared to the Cu based compounds at the same number
of valence electrons. Moreover, some of the Cu or Ni based compounds
are not ferromagnetic independent of the number of valence electrons
(not included in Fig.\ref{fig1}). Besides Mn$_2$VAl and Ir$_2$MnAl,
only compounds containing both, Fe and Mn, were found to exhibit HMF
character with magnetic moments according to the Slater-Pauling
rule.

\begin{figure}
\centering
\includegraphics[width=2.8in]{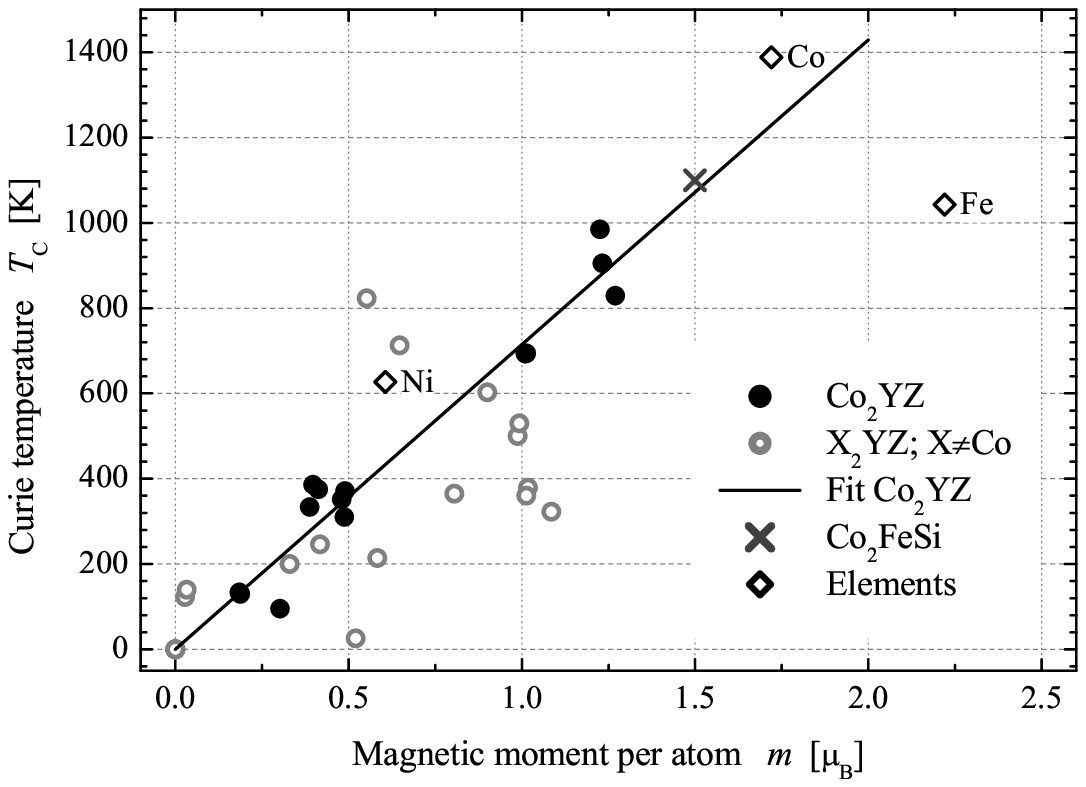}
\caption{Curie-temperatures of X$_2$YZ Heusler compounds.\\
         The line is found from a linear fit of the measured $T_C$
         for Co$_2$-based compounds (full dots).
         The elemental metals Fe, Co, and Ni are given for comparison.}
\label{fig_2}
\end{figure}

Plotting the Curie temperatures ($T_C$) of the known, $3d$ metal
based Heusler compounds as function of their magnetic moment results
seemingly in a linear dependence for Co$_2$YZ half-metallic Heusler
compounds (see Fig.\ref{fig_2}). According to this plot, $T_C$ is
highest for those half-metallic compounds that exhibit a large
magnetic moment, or equivalent for those with a high valence
electron concentration when comparing to the Slater-Pauling rule. By
extrapolating a linear dependence, $T_C$ is estimated to be above
1000K in compounds with $6\mu_ B$, that is with 30 valence electrons
per unit cell.

The origin of the seeming linear dependence is not clear at first
sight. In the molecular field approach, the Curie temperature of a
system with two magnetic sub-lattices (Co, Y) will be given by:
\begin{equation}
       T_C = \frac{1}{2} \max \left( T_{Co} + T_{Y} \pm \sqrt{(T_{Co} - T_{Y})^2 + 4T^2_{Co,Y}} \right)
\label{eq3}
\end{equation}
with the sub-lattice temperatures \cite{vvl45}
\begin{equation}
       \frac{3}{2} k_B T_i \propto J_i S_i(S_i+1).
\label{eq4}
\end{equation}
The $T_i$ are proportional to the Heisenberg exchange integral $J_i$
at site $i=Co,Y$. $S_i$ is the accompanied spin moment. $T_{Co,Y}
\propto J_{Co,Y}$ is a composite temperature. The $J_i$ describe the
interaction of the atom at site $i$ with the hole crystal.
$J_{Co,Y}$ is the pair interaction parameter between Co and Y sites
(for details see \cite{And63}).

The calculations revealed that the magnetic moments at Co and Y
sites increase simultaneously with increasing $n_V$. Thus the
expected non-linearity with $m$ has to be compensated by variation
of the Heisenberg exchange to result in the nearly linear dependence
on $n_V$. 

$T_C\approx\max(T_{Co},T_Y)$ is governed by the higher of the two
sub-lattice Curie temperatures if $T_{Co,Y}$ is small with respect
to $T_{Co}$ and $T_Y$. For the Co$_2$YZ compounds this is $T_Y$
as was also found for other Heusler compounds \cite{kue84}. Using
this restriction, the calculation yielded $T_C=1120$K for
Co$_2$FeSi, in good agreement with the experiment.

As a practical test for the models given here, we revisited
Co$_2$FeSi. This compound was previously reported to have a magnetic
moment of $5.9\mu_B$ per unit cell at 10K and a Curie temperature of
$>980$K \cite{nbr77}, whereas band structure calculations predicted
only $5.27\mu_B$ \cite{gdp02}. One expects, however, $M=6\mu_B$ for
the spin moment and $T_C$ to be clearly above $1000$K, from the
estimate given above. Polycrystalline Co$_2$FeSi samples were
investigated \cite{wfk05} and X-ray diffraction confirmed the $L2_1$
structure with a lattice parameter of $a=5.64$\AA. Low temperature
magnetometry gave a magnetic moment of $5.97\mu_B$ per unit cell at
5K, in excellent agreement to the Slater-Pauling rule. The Curie
temperature was found to be $(1100\pm20)$K. This value fits very
well the nearly linear behavior shown in Fig.\ref{fig_2}. All
experimental findings are supported by recent band structure
calculations revealing a half-metallic ferromagnet with a magnetic
moment of $6\mu_ B$, if using appropriate parameters in the self
consistent field calculations \cite{kff05}.

\section{Conclusion}
In summary, it was shown how the Heusler compounds can be described
in terms of the Slater-Pauling rule. This is particularly the case
for Co$_2$YZ compounds that exhibit half-metallic ferromagnetism.
Further, it was found that the Co$_2$YZ compounds exhibit a
seemingly linear dependence of the Curie temperature on the magnetic
moment. Using the practical example of Co$_2$FeSi, it turned out
that the given rule of thumb works for both magnetic moment and
$T_C$. The rules may be applied not only to ternary but also to
quaternary compounds like Co$_2$Y$_{1-x}$Y'$_x$Z if stabilized in
the $L2_1$ phase. Recent examples are Co$_2$Cr$_{1-x}$Fe$_x$Z
(Z=Al,Ga) \cite{bfj03,kuk04,kuf05}. An experimental challenge will
be to find Heusler compounds with magnetic moments above 6$\mu_B$
and to prove whether it is possible to find even higher $T_C$ in
this class of materials. Co$_3$Z compounds crystallize in a
hexagonal structure, unfortunately. Thus, one way to reach that goal
may be to stabilize Co$_{2+x}$Fe$_{1-x}$Z in the $L2_1$ phase.

\begin{acknowledgments}
This work is financially supported by the DFG (FG 559).
\end{acknowledgments}

\bibliography{MMM_DS04_fecher}

\end{document}